# Phase-ordering of charge density waves traced by ultrafast low-energy electron diffraction

**S. Vogelgesang[1], G. Storeck[1], J. G. Horstmann[1], T. Diekmann[1], M. Sivis[1], S. Schramm[1], K. Rossnagel[2], S. Schäfer[1], C. Ropers[1,3,*]**

[1]*University of Göttingen, IV. Physical Institute - Solids and Nanostructures, Germany*
[2]*Institute for Experimental and Applied Physics, University of Kiel, Germany*
[3]*International Center for Advanced Studies of Energy Conversion (ICASEC), University of Göttingen, Germany*
**Email: claus.ropers@uni-goettingen.de*

**We introduce ultrafast low-energy electron diffraction (ULEED) in backscattering for the study of structural dynamics at surfaces. Using a tip-based source of ultrashort electron pulses, we investigate the optically-driven transition between charge-density wave phases at the surface of 1T-$TaS_2$. The large transfer width of the instrument allows us to employ spot-profile analysis, resolving the phase-ordering kinetics in the nascent incommensurate charge-density wave phase. We observe a coarsening that follows a power-law scaling of the correlation length, driven by the annihilation of dislocation-type topological defects of the charge-ordered lattice. Our work opens up the study of a wide class of structural transitions and ordering phenomena at surfaces and in low-dimensional systems.**

The reduced dimensionality and broken symmetry of a surface endows it with unique physical and chemical properties that drastically differ from the bulk[1,2]. Prominent surface-specific features involve the electronic, atomic and spin structure, as manifest in modified band structures[3,4], surface reconstructions[1] or topological states[2,5]. Many of these phenomena exhibit highly-complex couplings and correlations, which are difficult to disentangle using steady-state analyses of systems in equilibrium. As a result, ultrafast spectroscopy has become an indispensable means to identify the hierarchy and strengths of interactions in the time-domain, by probing the response of materials and surfaces excited strongly out of equilibrium[6]. Specifically, time-resolved realizations of optical and photoemission spectroscopy yield comprehensive insights into the transient state of the electron and spin systems[3,5,7–11]. In contrast, access to the structural degrees of freedom with ultimate surface sensitivity and high temporal resolution remains limited, despite notable achievements in time-resolved reflection high-energy electron diffraction (RHEED)[12–16]. In order to reach a detailed and quantitative understanding of ultrafast structural dynamics at surfaces, a time-resolved implementation of low-energy electron diffraction (LEED) is highly desirable. Although LEED is the most widely used and broadly applicable diffractive technique for surface characterization, an ultrafast realization has proven very challenging[17–21]. Recently, using the monolayer sensitivity of low-energy electrons, we introduced ultrafast LEED in transmission, studying the dynamics of a polymer superstructure on freestanding graphene[20]. However, a backscattering geometry promises a greatly expanded range of accessible systems and phenomena, including the dynamics of surface reconstructions, molecular adsorbates, or structural phase transitions. Moreover, simultaneously high temporal and momentum resolution may provide for unprecedented insights into the kinetics of phase-ordering, for example in correlated materials.

In this work, we present the development of ultrafast low-energy electron diffraction (ULEED) and demonstrate its applicability for the study of structural phase transitions and the resulting phase-ordering at surfaces. In particular, we investigate the optically-driven transition between two prominent charge-density wave (CDW) phases at the surface of single crystalline 1T-$TaS_2$. We track the formation and non-equilibrium temporal evolution of the incommensurate CDW phase, and identify a coarsening of the CDW texture by analyzing diffraction intensities and spot profiles. Enabled by compact electron sources based on nanotip photoemitters, ULEED represents a powerful and complementary addition to the toolbox of ultrafast surface science.

ULEED is part of a larger family of optical-pump/electron-probe schemes, in which an ultrashort electron pulse samples the momentary state of an optically-excited system by diffraction. In these approaches, the temporal resolution is limited by the electron pulse duration at the specimen position, which is broadened by Coulomb interactions within the electron pulse, velocity dispersion and path length differences upon propagation from the photoelectron source to the sample. In high-energy ultrafast electron diffraction (UED)[22–28] and ultrafast transmission electron microscopy (UTEM)[29–33], femtosecond temporal resolution is achieved by radio-frequency pulse compression[34–36] or tailored gun designs[22,32,37]. Due to longer electron flight durations, time-resolved experiments with low-energy electron pulses face the difficulty of a greatly increased impact of any effect leading to pulse broadening. Moreover, in order to minimize the sample-source distance in the backscattering geometry of LEED, the outer diameter of a pulsed electron gun needs to be reduced accordingly for obtaining diffraction images while avoiding shadowing (cf. Fig. 1a).

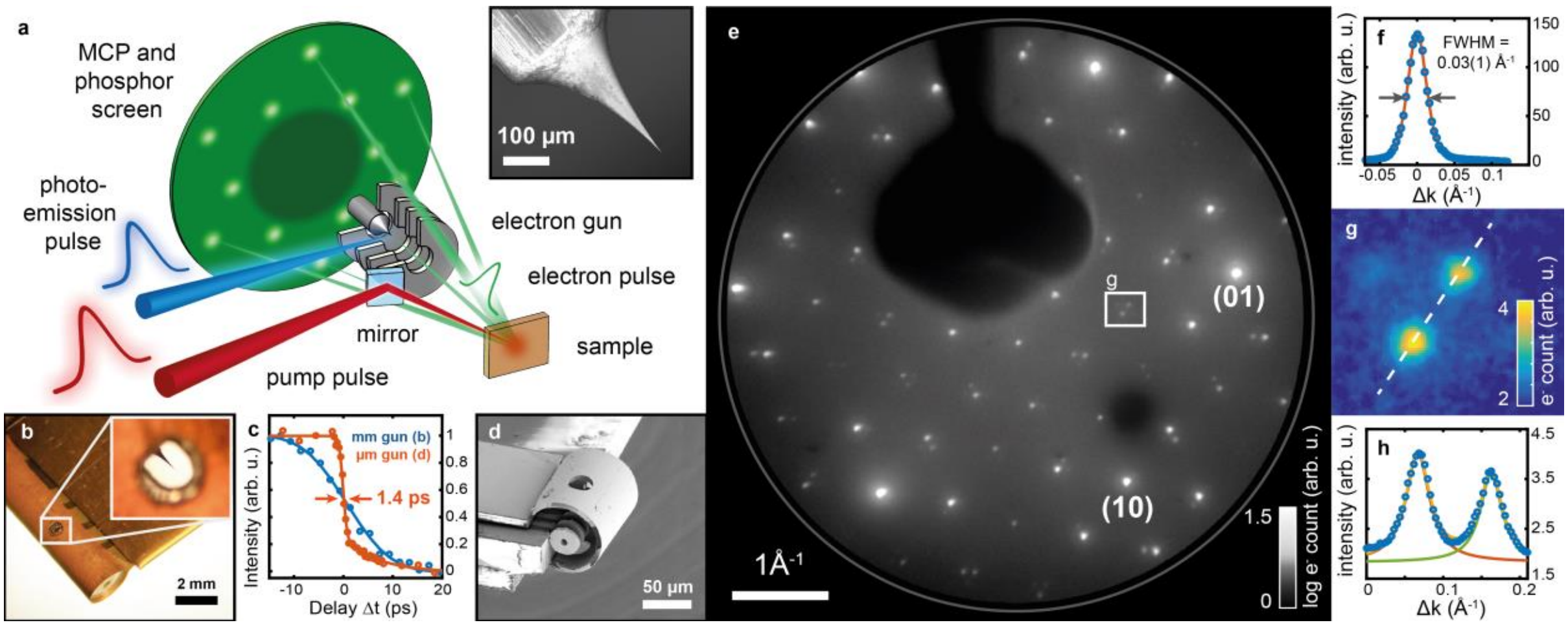

**Figure 1 | ULEED setup and high-resolution diffraction pattern from 1T-$TaS_2$. a,** Schematic of the experimental setup for ultrafast low-energy electron diffraction (ULEED). Inset: Electron micrograph of an electrochemically etched tungsten tip used as a photoemitter. **b,** mm-sized laser-driven electron gun. Inset: Tungsten tip, visible through the hole for laser illumination. **c,** Electron pulse durations of mm-sized electron gun (16.4 ps (FWHM) at 100 eV) and µm-sized electron gun (1.4 ps (FWHM) at 50 eV), determined by fast component of NCP diffraction peak suppression. **d,** µm-sized electron gun, prepared using nanofabrication techniques[38]. **e,** LEED pattern of the NC CDW room temperature phase, recorded with pulsed 100-eV electrons from the mm-sized electron gun (logarithmic color scale). A retarding voltage of -20 V is applied at the detector front plate. **f,** Line profile of the $(1\bar{1})$ diffraction peak, illustrating high transversal coherence of the source. The fitted spot width of 0.03 Å$^{-1}$ (FWHM) corresponds to a transfer width of 21 nm. **g,** Close-up of region marked in (e), showing second-order CDW diffraction spots. **h,** Line profile of CDW diffraction spots shown in (g), fitted with Lorentzian peak profiles.

Addressing these issues, we developed two particularly compact ultrafast low-energy electron source designs at millimeter and micrometer scales, respectively, each composed of a sharp tungsten tip (apex radius <25 nm) inserted into an electrostatic lens assembly for acceleration and beam focusing. Details of the electron gun designs and their characteristics are described in the methods and in Ref. [38]. We generate ultrashort electron pulses via two-photon photoemission (2PPE) by illuminating the tip with 400 nm laser pulses (40 fs duration), as recently demonstrated for low-energy transmission experiments[20] and in UTEM[32]. Beam collimation at typical operation energies (40-150 eV) is ensured by optimization of all corresponding electrode voltages. The temporal resolution of the two setups is characterized by electron-laser cross-correlation using the transient-electric-field (TEF) effect[20,39–41] or the fastest structural responses in backscattering diffraction (cf. Fig. 1c). We obtain a temporal resolution down to 16 ps for the mm-sized electron gun (100 eV energy) and 1.4 ps for the micrometer-sized gun (50 eV), sufficient for investigating a variety of structural evolutions at surfaces. Nanometric photocathodes constitute electron sources with a strongly confined emission area[42,43], leading to beams of high transversal coherence length[20,32,44]. Employing high-dynamic range detection using a phosphor-screen microchannel plate (MCP, Hamamatsu F2226-24P) detector and a cooled CMOS camera, we obtain LEED images of excellent quality, with a momentum resolution $\Delta k_s = 0.03$ Å$^{-1}$ (cf. Fig. 1f, recorded with mm-sized gun) corresponding to a transfer width of $2\pi/\Delta k_s = 21$ nm for a spot size on the sample below 100 µm (full-width-at-half-maximum, FWHM).

In a first application of these experimental capabilities, we study the dynamics of a structural phase transition at the surface of 1T-$TaS_2$. This compound exhibits a variety of equilibrium[45,46] and metastable[47] CDW phases that are coupled to periodic lattice distortions (PLD) and, at low temperatures, are accompanied by electron localization[48] or orbital order[49]. The room-temperature, so-called "nearly commensurate" (NC) CDW phase, features a particularly interesting structure: It is composed of a close-to-hexagonal arrangement of domain-like commensurate (C) areas separated by a network of discommensurations, which lacks complete periodicity[50,51]. A LEED pattern of the 1T-$TaS_2$ surface in the NC phase, cleaved in ultrahigh vacuum, is displayed in Fig. 1e (100 eV energy, angle of incidence 6°, logarithmic intensity scale, recorded using nanotip photoelectrons from the mm-sized electron gun). The image exhibits a multitude of sharp and well-separated diffraction peaks spanning three orders of magnitude in intensity. Specifically, the atomic-lattice Bragg peaks (indexed, for simplicity hereafter called Bragg peaks) are surrounded by six PLD-induced satellite spots each, which are rotated by an angle of ~12° to the lattice[45]. The large transfer width of the setup and the high signal-to-noise ratio allow us to clearly resolve the closely-spaced higher-order diffraction peaks (Figs. 1g, h), which result from the domain-like structure of the NC phase[50]. At temperatures above 353 K, 1T-$TaS_2$ exhibits a transition to an incommensurate (IC) CDW phase with wave vectors parallel to those of the atomic lattice[45]. Hence, this structural phase transition is associated with the appearance of satellite diffraction spots between the lattice Bragg peaks (Figs. 2b,c), as recently demonstrated in UED at high electron energies in transmission through a bulk film[26,28,52].

We now employ ULEED to examine this NC-to-IC transition at the surface, beginning with results from the mm-sized electron gun. The structural dynamics is triggered by optical pump pulses of 200 fs duration and a center wavelength of 1030 nm. A repetition rate of 25 kHz was selected to ensure structural and thermal relaxation between consecutive pump pulses. (A complete relaxation of the structure was found to span tens to hundreds of ns, depending on pump fluence, as measured with two electronically synchronized laser systems.) To provide a homogeneous sample excitation across the electron beam (100 µm FWHM), the pump beam is focused to a diameter of ~300 µm (FWHM) on the sample, with incident fluences ranging from 0.56 mJ/cm² to 5.65 mJ/cm². The transient state of the sample's surface structure is subsequently probed by electron pulses with a kinetic energy of 100 eV after a variable delay time $\Delta t$. Figure 2b displays the diffraction pattern of the sample after optical excitation to the IC phase ($\Delta t$ >734 ps), with CDW diffraction peaks in-line with the reciprocal vectors of the atomic lattice. Due to the weak and harmonic PLD, higher-order CDW diffraction peaks are practically absent in the IC phase[53]. A difference image (Fig. 2c) and lineouts (Fig. 2d) illustrate the suppression of the NC phase (blue) and the appearance of the rotated IC spots (red), together with an increase of the diffuse background.

For all time delays and ten different pump fluences, we evaluate the integrated intensities of the Bragg, NC- and IC-CDW diffraction peaks, plotted in Fig. 2f. Normalized diffraction peak intensities at large delays ($\Delta t$ =1134 ps) as a function of fluence are shown in Fig. 2e. For fluences below 2.83 mJ/cm², the NC-CDW phase is transiently suppressed by up to about 25% (see blue time curves in Fig. 2f, middle) due to the pump-induced heating of the surface (Debye-Waller effect), recovering on a timescale of several nanoseconds via thermal diffusion into the bulk, consistent with the out-of-plane thermal conductivity of the material[54]. At a pump fluence of 3.4 mJ/cm², we observe a sharp threshold behavior, leading to a full suppression of the NC phase. Simultaneously, the IC phase appears, and its diffraction spot intensity levels off for higher fluences (with some indication of a thermal suppression at the highest fluence, cf. Fig. 2e). While the NC phase is rapidly suppressed (see also Refs. [26,28]), the IC spot intensity continuously increases over several hundred picoseconds. Measurements conducted with the µm-sized electron (Fig. 2g, intensity integrated over a disk-shaped area in k-space 0.12 Å$^{-1}$ wide) illustrate the slower growth of the IC intensity at higher temporal resolution. These observations cannot be accounted for by a mere thermal relaxation of the newly created IC phase, as evidenced by the slower nanosecond recovery of both the Bragg spot intensity and the thermal suppression of the NC spots below threshold (green and light blue curves in Fig. 2f, respectively).

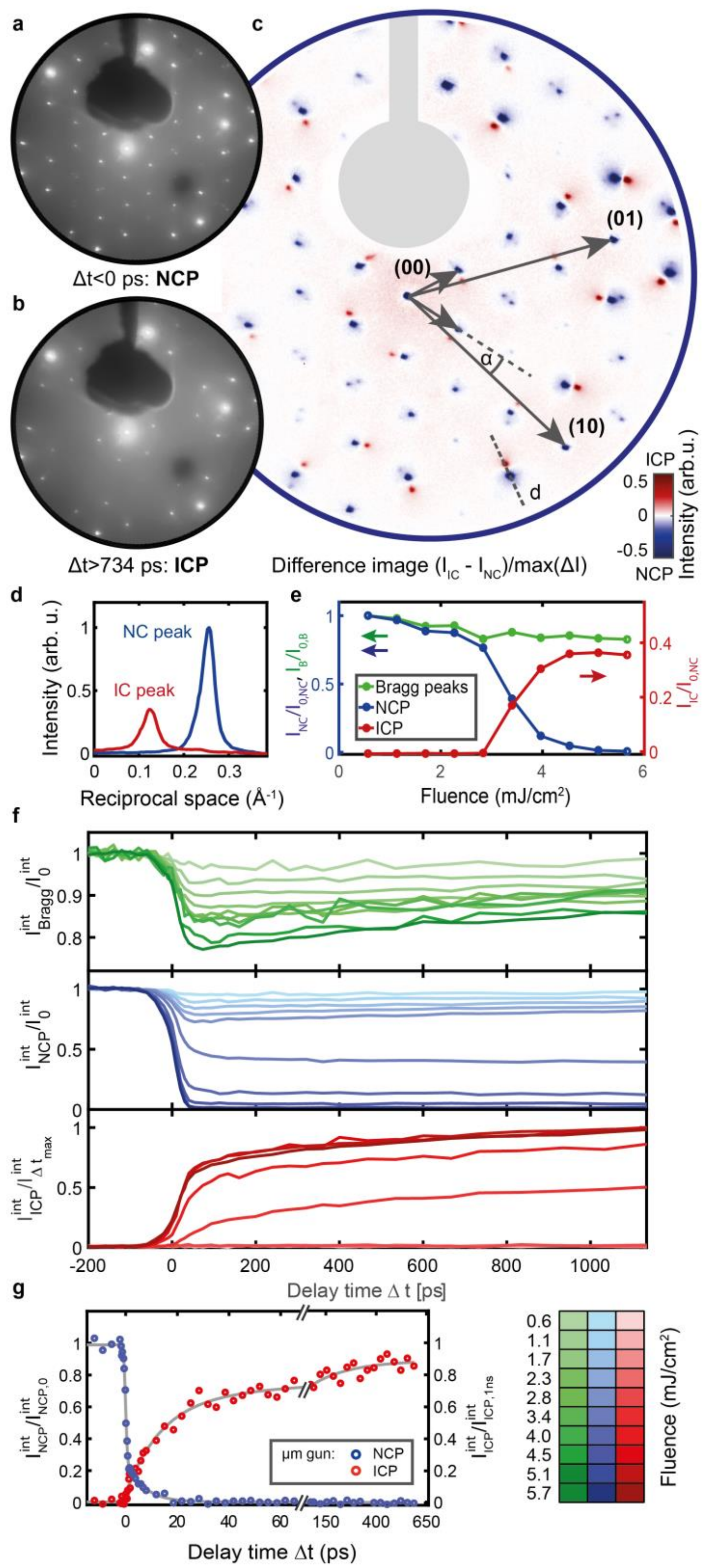

**Figure 2 | ULEED of the structural phase transition between CDW phases. a,** Diffraction pattern of the NC phase recorded at $\Delta t < 0$ ps. **b,** Diffraction pattern of the sample optically-pumped to the IC phase ($\Delta t > 734$ ps). Note the CDW satellite spots, which are rotated by -12° with respect to the NC phase in (a). **c,** Difference image of the pumped and unpumped diffraction patterns (a, blue) and (b, red). **d,** Line profile of NC and IC CDW diffraction spots marked in (c). **e,** Fluence dependence of the NC, IC and Bragg diffraction peak intensities at $\Delta t =$ 1.1 ns, normalized to the NC and Bragg spot intensities $I_{0,NC}$ and $I_{0,B}$ at lowest pump fluence and negative times. **f,** Integrated diffraction spot intensities of the Bragg (top), NC (center) and IC (bottom) diffraction peaks vs. delay time for ten optical pump fluences, normalized to the signal at negative times. **g,** Integrated CDW diffraction peak intensities recorded using the μm-sized gun.

One of the central questions in the transition between charge-ordered states of different translational symmetries concerns the microscopic mechanisms involved in the emergence of long-range order. As there is no homogeneous global deformation from the NC to the IC phase (as, e.g., in the martensitic transformation), the transition will involve substantial transient disorder. The high momentum resolution afforded by our short-pulsed electron beam allows us to identify the associated time-dependent changes in correlation length. Specifically, we find that the IC diffraction spots initially appear as rather broad peaks that significantly sharpen within few-hundred ps after the excitation from $w_{exp} = 0.06$ Å$^{-1}$ at $\Delta t = 20$ ps to $w_{exp} < 0.038$ Å$^{-1}$ for $\Delta t > 400$ ps, where $w_{exp}$ is the measured diffraction spot width (FWHM, taken as the geometric mean of both semi-axes).

A quantitative analysis of the spot profiles taking into account the instrument response function (see Methods for details) reveals a clear monotonous peak narrowing (Figs. 3a, b), which, in terms of the real-space structure, corresponds to a growth of the correlation length $L_{IC}$ with time $\Delta t$. We observe diffraction peaks which are slightly elongated in the azimuthal direction (cf. Fig. 3a,b). Plotting the determined correlation length $L_{IC}$ over time on a double-logarithmic scale, we observe an approximate power-law scaling $L_{IC} \propto \Delta t^{x}$ with $0.51 \pm 0.03$ (Fig. 3c, inset). Interestingly, we find the growth exponent to be universal for this transition in the sense that it does not depend on the excitation fluence in the range studied.

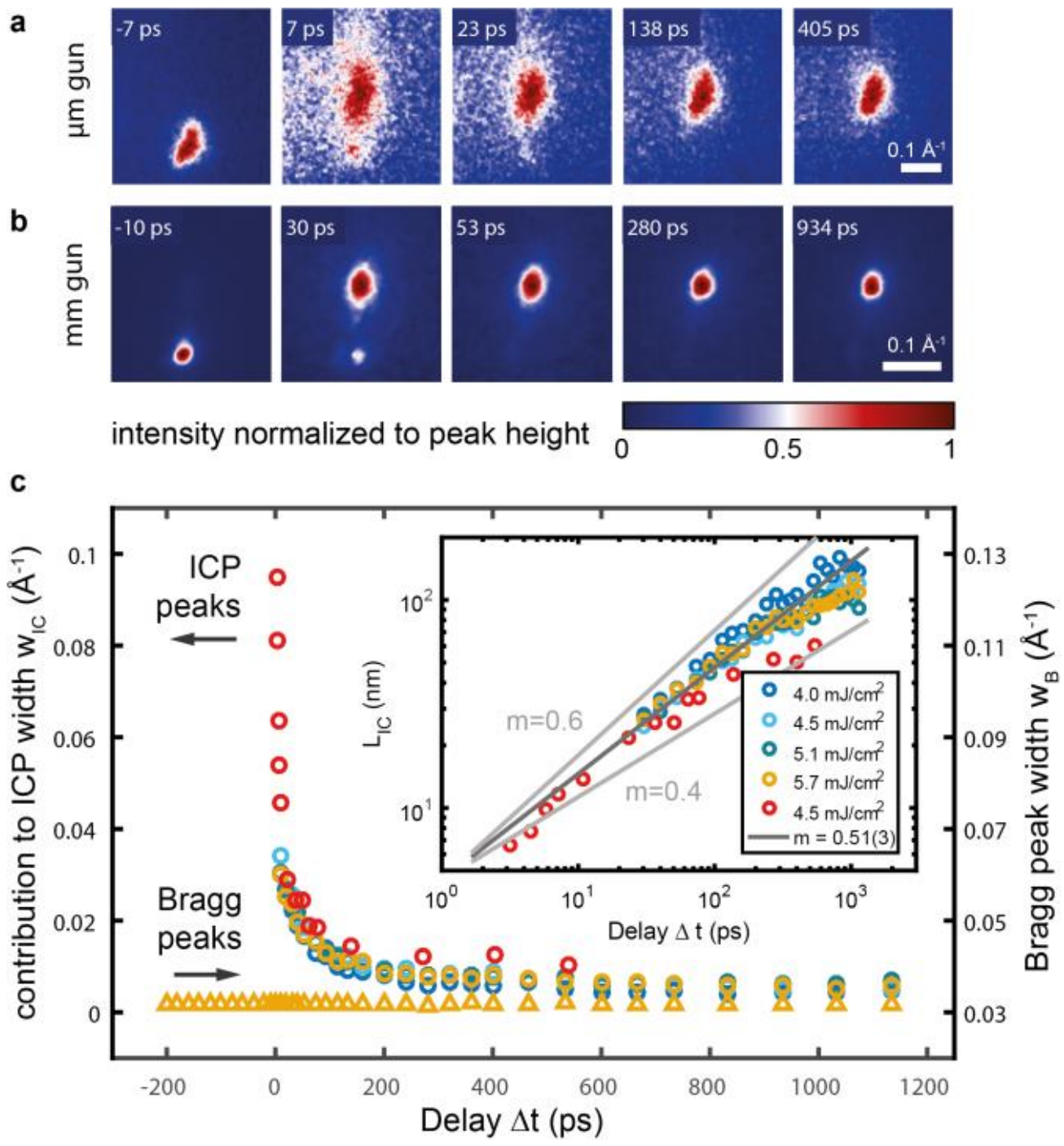

**Figure 3 | Phase-ordering kinetics of IC CDW governed by topological defects. a,** CDW diffraction peaks during NC-IC-phase transition, illustrating the ICP diffraction narrowing, (data recorded with µm-sized gun, intensity corrected for linear background, color scale normalized to peak height). **b,** ICP diffraction peak narrowing recorded with mm-sized gun. **c,** Time-dependent contribution to the width (FWHM) of the ICP diffraction peaks and width of lattice diffraction peaks for five optical pump fluences. The decreasing contribution to the peak width is related to an increasing correlation length of the structure. Inset: A double-logarithmic plot of the correlation length corrected for the instrument response function indicating a scaling law for the phase-ordering kinetics. Red circles: Data recorded with µm-sized gun.

Generally, the characteristic length scale of a system undergoing coarsening or phase-ordering kinetics typically follows a power-law scaling over time, i.e. $L \propto t^{x}$. The growth exponent $x$ depends on the type of order parameter (conserved/non-conserved)[55], the effective dimensionality[56] and in-plane anisotropy, and, in the case of defects, their diffusion[56] and interaction forces[57]. A value of $x = 0.5$ is found in cases of non-conserved order parameters, as well as specific scenarios such as pair annihilation at homogeneous defect densities[55,56].

A decrease in spot width for selected delay times was very recently observed by UED[28] and ultrafast X-ray diffraction[58], but was not closely traced over time. Specifically, Ref.[28] hypothesized that the growth may be initiated at the NC discommensurations, while in Ref.[58], coarsening of the IC phase was attributed to the growth of well-defined domains enclosed by linear domain walls. However, so far, a detailed microscopic picture for this phase ordering is still missing, which calls for a time-dependent description relating to theoretical treatments of the phase transitions in these materials[51,53,59,60].

Before discussing the physical mechanism underlying the coarsening behavior observed, it is instrumental to consider the nature of the order parameter defining the incommensurate charge density wave system and the contributions to the free energy governing its phase-ordering kinetics. The order parameter of the system is the charge density modulation $\alpha(\vec{r})$, which exhibits a hexagonal structure given by the superposition of three modulations along the directions of the CDW wavectors $q_j$: $\alpha(\vec{r}) = \mathrm{Re}\big(\psi_1(\vec{r}) + \psi_2(\vec{r}) + \psi_3(\vec{r})\big)$ [59]. The three components $\psi_j(\vec{r}), j = 1,2,3$ of the order parameter are each composed of a plane wave contribution and a spatially varying complex amplitude $\phi_j(\vec{r})$, namely $\psi_j(\vec{r}) = \phi_j(\vec{r}) \exp\big(i\vec{q}_j \cdot \vec{r}\big)$. Generally, this description of the order parameter applies to both C and IC phases. However, it is important to realize that there are notable differences between incommensurate and commensurate CDWs regarding their respective texture or domain structure. Specifically, in commensurate superstructures, a discrete set of degenerate ground states exists, distinguished by integer shifts of the real-space lattice vectors, which leads to the possibility of stabilizing well-defined domains enclosed by domain walls. Incommensurate systems, on the other hand, support either staircase-type (an example is the NCP) or "floating" phases, such as the ICP, which can be continuously translated across the atomic lattice without being locked to it [61]. Following the seminal works of Overhauser and McMillan, the elementary excitations of the order parameter in this latter type of IC phase are well-established as phonon-like phase modulations (also referred to as "phasons" [59,62,63]) and dislocation defects[59,60,64]. Whereas phasons represent wavelength modulations of the CDW, the dislocations are defects of a topological character, which exhibit phase singularities in two of the three amplitudes $\phi_j(\vec{r})$, in the form of a vortex-like phase spiral (see insets of Fig. 4a-d)[59]. In a diffraction experiment as conducted here, both types of excitations can be identified. Phasons lead to a suppression of diffraction peak intensity (Debye-Waller effect; more precisely, a suppression that depends on the phason spectrum)[62]. The vortex-like dislocations, on the other hand, limit the correlation length of the structure and therefore induce a peak broadening.

In order to obtain a qualitative understanding of the charge-ordering kinetics and the role of these defects in this incommensurate system, we perform model computations within a time-dependent Ginzburg-Landau approach. Specifically, we solve a set of coupled equations of motion for the order parameters $\psi_j(\vec{r})$[55],

$$\frac{\delta\psi_j}{\delta t} = -\frac{\delta F_{IC}[\psi_1, \psi_2, \psi_3]}{\delta\psi_j^*},$$

with a phenomenological free-energy functional $F_{IC}[\psi_1, \psi_2, \psi_3]$ containing essential terms (kinetic energy, nonlinearity, phasing term) previously invoked in theoretical descriptions of this material[51,59]. Thus, the simulations track the evolution of the charge order as it is governed by the relaxation of the free energy in the IC phase. As initial conditions for the computations, we use the initial charge of the NC phase with some random phase and amplitude noise (see details in Methods). Figure 4 displays an exemplary sequence of frames from these simulations, showing the spatial charge order, together with the resulting predicted diffraction spot profiles (insets; the squared Fourier magnitudes of the CDW yield the spot profiles for small displacements) and distributions of individual components $\psi_i(\vec{r})$ of the order parameter.

These frames illustrate the emergence of long-range order in the IC phase as a result of the global transformation from the NC state that exhibits different symmetries and wavevectors (orientation and magnitude). As this transformation has stochastic elements and is locally driven in effectively decoupled nanoscale regions, it represents a quench: It necessarily involves the creation of a highly disordered state that locally exhibits IC periodicity and orientation, but which is globally characterized by a considerable density of fluctuation modes and dislocation-type defects and, thus, lacks long-range phase coherence.

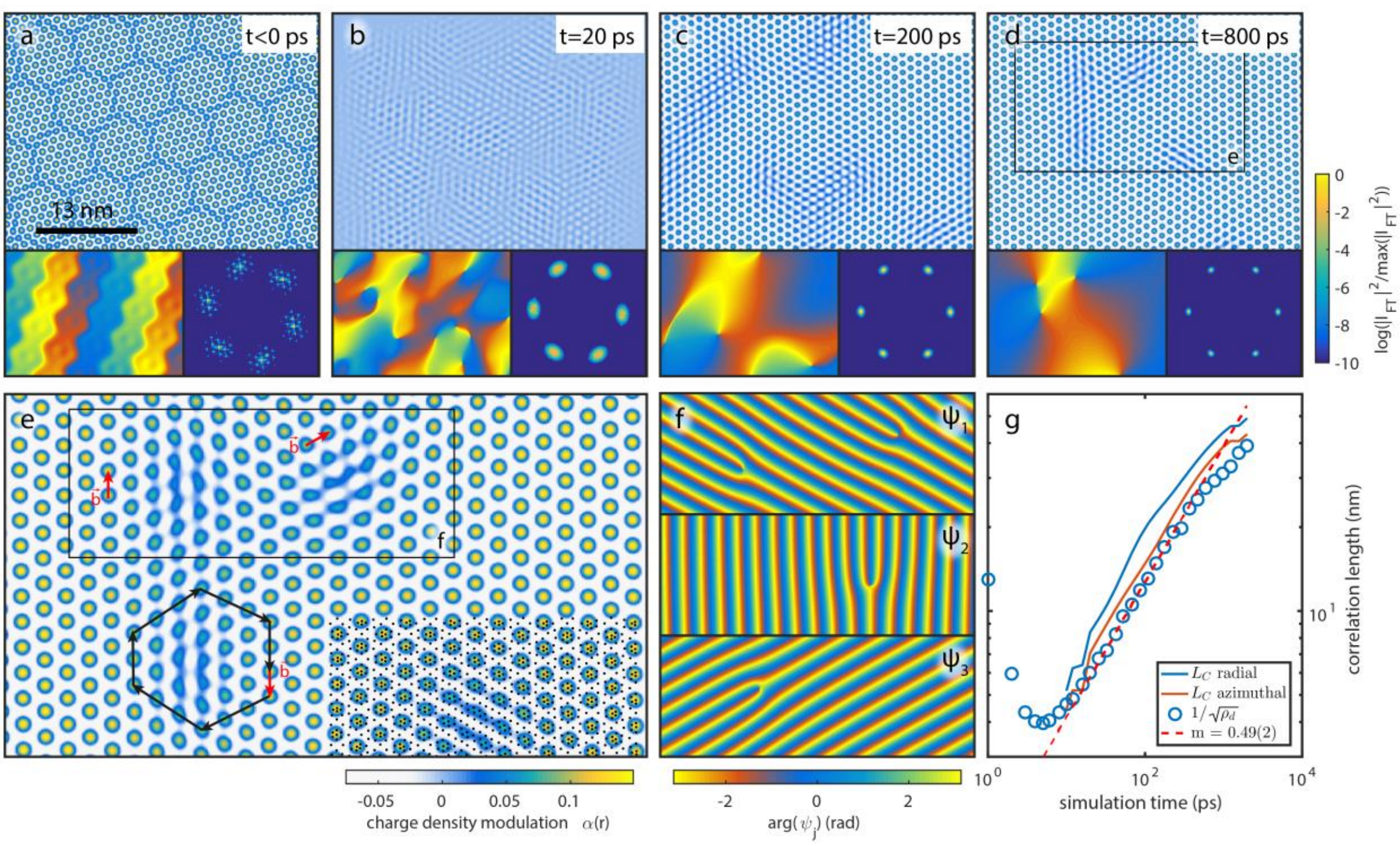

**Figure 4 | Numerical simulation of phase-ordering kinetics. a-d,** Emergence of a long-range-ordered IC phase associated with the annihilation of topological defects, starting from the NC phase (a). Bottom left insets show the phase variations given by $\arg[\psi_1(\vec{r})]$, removing the oscillatory components within the CDW unit cell. The different frames (b-d) exhibit the phase singularities at topological defects. Bottom right insets: Normalized Fourier magnitudes of $\alpha(\vec{r})$, illustrating the time-dependent narrowing of the diffraction peaks. **e**, Close-up of a region (marked in d) displaying several topological defects. Each defect is characterized by a Burgers vector (red arrow) pointing in direction of $\pm\vec{q}_j$. Lower right corner: PLD induced by the CDW with the atomic displacements exaggerated eight times. **f**, Phases of $\psi_{1,2,3}(\vec{r})$ shown for region marked in e, illustrating that each defect is composed of dislocations in two of the three contributions to the order parameter. **g**, Temporal evolution of the correlation length, obtained from the FWHMs of the peaks in the Fourier magnitude, together with mean defect distance (circles). A linear fit to $1/\sqrt{\rho_d}$ yields a growth exponent of $\mathrm{m} = 0.49 \pm 0.02$ (dashed line). The initial drop in average defect distance at very early times signifies the initial formation of dislocations from the defect-free NC state.

The characteristic behavior of the phase ordering kinetics is evident from the simulations. While strong phase gradients in the complex amplitudes $\phi_j(\vec{r})$ (composed of phasons) continuously decay to minimize the free energy[64], the topological defects, evident as phase singularities (Fig. 4f and insets in 4a-d), are more persistent. In analogy to atomic lattices, the CDW dislocations are inherently stable entities characterized by a topological invariant, the Burger's vector (cf. Fig. 4e), pointing in one of the six possible directions given by the CDW. The model demonstrates that long-range order in this system is established by the decay of the dislocation density by pair annihilation. Generally, two processes are possible: Dislocations with opposite Burger's vectors may completely annihilate (number of defects reduced by two for every event), or two defects with an angle of 120° between the Burger's vectors may recombine into a new defect described by their vector sum.

The phase-ordering can now be analyzed by studying the time-dependence of the defect density (Fig. 4g), or equivalently, the correlation length of the simulated charge-ordered state. In agreement with the experiment, the model yields a power-law coarsening with an exponent $x = 0.5$ for early times and, interestingly, even predicts the observed moderate slowing at later times. Thus, combining our experimental analysis with these simulations strongly suggests that the growth in correlation length is caused by the natural process of defect annihilation. Closer inspection reveals that in the simulations, the slowing of the coarsening coincides with the pairing of partial dislocations[61], i.e., dislocations in only one of the order parameter components $\psi_j(\vec{r})$, which are initially not bound in the form predicted by McMillan[59]. Whether or not this mechanism is indeed responsible for the reduction of power exponent indicated by the experiment, cannot be finally determined at this point, since other effects such as pinning of the CDW at structural defects may be equally responsible. Also, diffusion-limited particle-antiparticle annihilation in two dimensions ($x = 0.25$) may become a relevant description at later times[56].

In conclusion, introducing an ultrafast surface-sensitive structural probe, namely ultrafast low-energy electron diffraction, our work traces the phase-ordering kinetics of a charge-density wave system effectively quenched by optical excitation. Generally, our findings of kinetics driven by dislocation defects represent a striking counterpart of ultrafast phason dynamics in the absence of topological defects, as recently observed in charge- and spin-ordered nickelates[63]. In fact, the present system may exhibit both phenomena, and the extent to which the decay of non-topological phase fluctuations contribute to the relaxation will be a subject of further study. Similarly, the initial concentration of topological defects and its dependence on the spatial textures of both phases involved remains an intriguing issue. The study illustrates the use of spot-profile analysis in ultrafast electron diffraction, which will be applicable to a wide range of ordering phenomena in solids and at surfaces.

**Acknowledgements**

This work was funded by the European Union (EU) within the Horizon 2020 ERC-StG. "ULEED" (ID: 639119). We gratefully acknowledge insightful discussions with S. V. Yalunin and A. Zippelius.

## Methods:

### Analysis of diffraction peak intensities and profiles

To obtain the integrated intensities of both the lattice and CDW diffraction peaks from the raw data, a circular area of interest (radius $r$) around each spot of the diffraction pattern is summed up after background subtraction. The latter is determined within a ring of width $dr$ at the edge of the area of interest. For the measurements conducted with the mm-sized and μm-sized electron gun, the employed values of $r$ are 0.058 $Å^{-1}$ (20 pixels) and 0.066 $Å^{-1}$ (30 pixels), respectively. The width $dr$ is 3 pixels for both cases.

In order to determine the time-dependent contribution of the IC phase to the CDW diffraction peak widths, we perform a spot profile analysis of the ICP diffraction peaks via two-dimensional fits to the data, considering the instrument response function. Specifically, at each time delay, the 2d Cauchy distribution

$$C(x,y) = \frac{1}{2\pi\sigma_1\sigma_2} \cdot \left(1 + \left(\frac{x}{\sigma_1}\right)^2 + \left(\frac{y}{\sigma_2}\right)^2\right)^{-\frac{3}{2}}$$

is convoluted with a numerical representation of the instrument response function, and then fitted to the ICP diffraction peaks including a linear background. For the datasets recorded with the mm-sized gun and the μm-sized gun, the instrument function is approximated by a background-corrected and normalized area of interest around the (01)-Bragg peak of the atomic lattice and the sharpest corresponding NC CDW peak, respectively. During the fit procedure, $\sigma_1$ and $\sigma_2$ represent the two half axes of the elliptical Lorentzian profile in the radial and azimuthal directions with respect to the associated Bragg peak. The full-width-at-half-maximum in both directions can be obtained via $w_{IC} = 2\sigma\sqrt{2^{2/3} - 1}$.

In Fig. 3c, the contribution to the width of the ICP peaks $w_{IC}$ is compared to the width of the atomic lattice Bragg peaks $w_B$ (right axis). For this purpose, the Bragg peaks are fitted with Voigt-Profiles, and their full-width-at-half-maximum is numerically determined.

### Ultrafast Low-Energy Electron Guns

#### mm-sized electron gun:

The laser-driven mm-sized electron source consists of four metal electrodes (2 mm outer diameter, aperture radius 200 μm), which act as a suppressor, an extractor and an electrostatic einzel lens in combination with a grounded aperture. To generate short electron pulses via two-photon-photoemission, a nanometric tungsten inserted into the stack of electrodes can be illuminated by fs laser pulses from the side through a 500 μm hole. Finally, the electrodes and the nanotip photocathode are contacted and shielded by a flexible printed circuit board (FPCB). Due to its small outer diameter, the FPCB housing allows for operational distances of few millimeters from the sample position, while maintaining the visibility of the diffraction pattern.

#### μm-sized electron gun:

In contrast to the mm-sized gun, for a laser-driven electron source on the μm-scale, assembly requires micromanipulation. Here, focused ion beam techniques are the essential tool to fabricate and align apertures with a diameter of 30 μm as well as a nanometric tungsten tip. The electrodes are contacted to a circuit board produced by optical lithography via Platinum deposition. Further details of the fabrication of the μm-electron design are given in Ref. [38]

## Modelling of CDW kinetics

To simulate the emergence of the ICP CDW from a disordered state, we employ a simplified version of the free energy used by McMillan[59] and Nakanishi[51,53] in a time-dependent Ginzburg-Landau approach[55]. The evolution of each contribution $\psi_j$ ($j = 1,2,3$) to the order parameter $\alpha(r)$ is given by

$$\frac{\delta\psi_j}{\delta t} = -\frac{\delta F_{IC}}{\delta\psi_j^*}$$

with

$$F_{IC} = \int d\vec{r} \left[ \sum_j a\left|\left(\vec{q}_j \cdot \nabla - i\vec{q}_j^{\,2}\right)\psi_j\right|^2 + b\left|\vec{q}_j \times \nabla\psi_j\right|^2 + c\left(\frac{2}{3}\left|\psi_j\right|^3 - g\left(1 - \frac{d}{c}\right)\left|\psi_j\right|^2\right) + d\left(\psi_1\psi_2\psi_3 + \psi_1^*\psi_2^*\psi_3^*\right) \right]$$

and

$$\frac{\delta F_{IC}}{\delta\psi_j^*} = -a\left(\vec{q}_j \cdot \nabla - i\vec{q}_j^{\,2}\right)^2\psi_j + b\left(\left(\vec{q}_j \cdot \nabla\right)^2 - \vec{q}_j^{\,2}\Delta\right)\psi_j + c\left(\left|\psi_j\right| - g\left(1 + \frac{d}{c}\right)\right)\psi_j + d\left(\psi_{j+1}^*\psi_{j+2}^*\right).$$

Here, the first and second term adjust $\psi_j$ to the periodicity and orientation given by $\vec{q}_j$, whereas the third term adjusts the amplitude of $\psi_j$ to yield the value $g$. The last term induces the phasing of the $\psi_j$ to add up to $2\pi n$, i.e., it leads to the characteristic CDW pattern with circularly shaped charge accumulations in the CDW lattice.
The amplitude $g$ of the charge density modulation is given by the first-order Bessel function via $g = J_1(\vec{K} \cdot \vec{A})$. [62] The coefficients of the different terms in the free energy determine the rates, at which the amplitude and relative phases of the $\psi_j$ are locally adjusted (nonlinear terms) and the global curvature of the phase fluctuations ('kinetic energy' term) decays. Their balance influences the energy, mobility and effective interaction forces of the topological defects. For a wide range of parameters, we find the same qualitative features of topological defect annihilation governing the phase-ordering. The parameters used in the simulation for Fig. 4 are $a = 6/q_1^4$ ps$^{-1}$, $b = a/2$, $c = 1$ ps$^{-1}$ and $d = -0.25$ ps$^{-1}$. The ratio of $a$ and $b$ determines the stiffness of the CDW against compression and rotation, respectively, and $a > b$ leads to azimuthally elongated diffraction peaks, as observed experimentally.
The evolution equation is numerically integrated using a Split-Step-Fourier-Method, in which the first two terms are propagated in momentum space, and the amplitude and phase terms are separately applied in real space. As initial conditions, we use the NC charge order superimposed with amplitude (50% peak-to-peak) and phase ($\pm\pi$ peak-to-peak) noise of a 2.4 nm correlation length, with the phase noise applied in the different components adding to zero. For the computation of the diffraction peaks, we average over 10 realizations with different initial conditions on a numerical domain size of 100 nm in width.